\documentclass[iop]{emulateapj-rtx4}
\usepackage{amsmath}

\submitted{Accepted to ApJ 2013 October 23}
\shortauthors{TOFFLEMIRE ET AL.}
\shorttitle{T PYXIDIS X-RAY OUTBURST}
\begin{document}

\title{X-Ray Grating Observations of Recurrent Nova T Pyxidis During The 2011
  Outburst}
\author{Benjamin M. Tofflemire\altaffilmark{1},
Marina Orio\altaffilmark{1}$^{,}$\altaffilmark{2},
Kim L. Page\altaffilmark{3},
Julian P. Osborne\altaffilmark{3},
Stefano Ciroi\altaffilmark{4},
Valentina Cracco\altaffilmark{4},
Francesco Di Mille\altaffilmark{4},
and Michael Maxwell\altaffilmark{5}
}

\altaffiltext{1}{Astronomy Department, University of Wisconsin, Madison, 475
  N. Charter St., WI 53711, USA}
\altaffiltext{2}{INAF-Osservatorio Astronomico di Padova, Vicolo
  dell'Osservatorio 5, I-35122 Padova, Italy}
\altaffiltext{3}{Department of Physics and Astronomy, University of Leicester,
  Leicester LE1 7RH, UK}
\altaffiltext{4}{Department of Physics and Astronomy, Padova University, Vicolo
  dell'Osservatorio 3,  I-35122 Padova, Italy}
\altaffiltext{5}{Jeremiah Horrocks Institute, University of Central
  Lancashire, Preston PR1 2HE, UK} 

\begin{abstract}
  The recurrent nova T Pyx was observed with the X-ray gratings of {\sl
    Chandra} and {\sl XMM-Newton}, 210 and 235 days, respectively, after the
  discovery of the 2011 April 14 outburst. The X-ray spectra show prominent
  emission lines of C, N, and O, with broadening corresponding to a full width
  at half maximum of $\sim$2000-3000 km s$^{-1}$, and line ratios consistent
  with high-density plasma in collisional ionization equilibrium.  On day 210
  we also measured soft X-ray continuum emission that appears to be consistent
  with a white dwarf (WD) atmosphere at a temperature $\sim$420,000 K,
  partially obscured by anisotropic, optically thick ejecta. The X-ray
  continuum emission is modulated with the photometric and spectroscopic
  period observed in quiescence. The continuum at day 235 indicated a WD
  atmosphere at a consistent effective temperature of 25 days earlier, but
  with a lower flux.  The effective temperature indicates a mass of $\sim$1
  M$_\odot$. The conclusion of partial WD obscuration is supported by the
  complex geometry of non-spherically-symmetric ejecta confirmed in recent
  optical spectra obtained with the Southern African Large Telescope ({\sl
    SALT}) in November and December of 2012. These spectra exhibited prominent
  [O III] nebular lines with velocity structures typical of bipolar ejecta.
\end{abstract}


\section{INTRODUCTION}
\label{intro}
T Pyxidis is a recurrent nova (RN) with recorded outbursts in 1890, 1902,
1920, 1944, 1966, and 2011. Although models and observations indicate that all
classical novae (CNe) may be recurrent in nature
\citep{1976ARAA..14..119R,1996MNRAS.281..192R}, RNe comprise a rare subset
whose outbursts reoccur on human timescales (i.e., recurrence times $<$100
yr). The mechanism powering novae has been established as CNO-cycle driven
thermonuclear runaway (TNR) on the surface of a white dwarf (WD) in an
accreting binary system. About a dozen RNe are known in the Galaxy with a
handful in the LMC and M31 \citep{2010ApJS..187..275S}.

Apart from the frequency of outbursts, RNe differ from CNe in their mass
accretion rates and peak luminosities. Models reproducing the short recurrence
times of RNe \citep[e.g.][]{2005ApJ...623..398Y} require the combination of a
massive WD and large mass accretion rates ($\dot{M}$). Massive WDs require
less accreted material to reach the critical pressure for TNR and with a large
$\dot{M}$, outburst events are predicted on decade timescales. The resultant
outbursts consequently have less fuel, are generally less luminous, and have
shorter decay times than CN \citep{1987Ap&SS.131..493W}.

Unlike several other RN, T Pyx is a system with little apparent secondary
evolution and hosts an M dwarf companion \citep{2008A&A...492..787S}. It
belongs to a small group of RN including also two other objects, IM Nor and CI
Aql, where outbursts develop more slowly than in most RNe (see
\citealt{2010AJ....140...34S}). Furthermore, its orbital period is suggested
to be below the ``period gap'' \citep{2010MNRAS.409..237U} that canonically
marks a regime of gravitational radiation driven angular momentum loss with
$\dot{M}$$\sim$10$^{-10}$ M$_{\sun}$ yr$^{-1}$ \citep{CAPS43:2008}. This
estimate is two orders of magnitude less than inferred from observations
\citep{1998PASP..110..380P} and much less than what is predicted for the short
recurrence time observed.

T Pyx is also an enigmatic object because the mass ejected in each outburst
seems to be of order 10$^{-5}$ M$_\odot$, too large to have been accumulated
in the 12-44 yr between nova events. It has been speculated that the WD mass
is eroded by a large amount in each eruption
\citep{2008A&A...492..787S}. These estimates may be inflated, however, given
the evidence for bipolar outflows \citep{2011A&A...534L..11C} and a low
filling factor of ejected material \citep[as little as
$\sim$3\%;][]{2011A&A...533L...8S}

\begin{deluxetable*}{cccccc}[!thb]
\tablewidth{0pt}
\tabletypesize{\scriptsize}
\tablecaption{T Pyxidis X-Ray and Optical Observations}
\tablehead{
  \colhead{Date} &
  \colhead{Observatory} &
  \colhead{Instrument} &
  \colhead{Exp. Time} &
  \colhead{Days After} &
  \colhead{Mean Count Rate (cts ks$^{-1}$)} \\
  \colhead{} &
  \colhead{} &
  \colhead{} &
  \colhead{(ks)} &
  \colhead{Outburst} &
  \colhead{0.25-1.24 keV}
}
\startdata
2011 Nov 03 &\textit{Chandra}&HRC-S/LETG&40.13&210&$89.2\pm4.1$\\
2011 Nov 28 &\textit{XMM-Newton}&RGS 1-2&30.92&235&$44.5\pm1.1$\\
                     &                  & EPIC-pn & & & $1099\pm8$\\
\hline
2012 Nov 22 & {\sl SALT} &RSS PG900& ... & 588 &...  \\
2012 Dec 06 & &RSS PG2300&... & 602 &... \\
2012 Dec 17 & &RSS PG900&... & 617 &... \\
\enddata
\label{tab1} 
\end{deluxetable*}

Most intriguing, is the increasing recurrence time. Before 1966, T Pyx had a
mean recurrence time of $\sim$20 yr and yet, remained in quiescence for over
44 yr till the 2011 outburst. A drop in the average B-band magnitude by
$\sim$2 mag in the last 120 yr \citep{2010ApJ...708..381S} has been argued to
represent a decreasing accretion rate to explain the recurrence trend. The
above characteristics beg the question of whether T Pyx is in a transient
evolutionary state.  Authors prior to the 2011 outburst speculated that T Pyx
was heading toward a dormant phase \citep{2010ApJ...708..381S}.

\begin{figure}[th]
  \centering
    \includegraphics[keepaspectratio=true,scale=0.35,angle=-90]
    {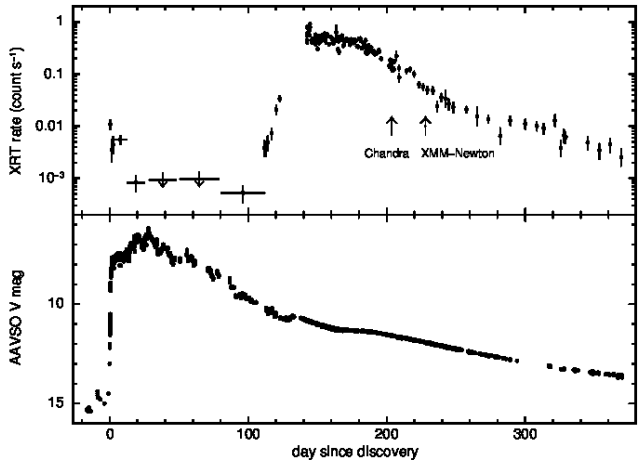}
    \caption{\textit{Top}: The {\sl Swift} 0.3-10 keV XRT PC mode grade 0-12
      light curve of T Pyx, shown with one bin per snapshot after $\sim$100
      days; at earlier times, some observations were combined to obtain
      detections or upper limits. The upper limits are calculated at 3
      $\sigma$ confidence according to the Bayesian method of
      \citet{1991ApJ...374..344K}. Arrows mark the dates of {\sl Chandra} and
      {\sl XMM} observations. \textit{Bottom}: V band CCD photometry from the
      AAVSO (http://www.aavso.org); only self-consistent data sets are shown.}
    \label{fig:slc}
\end{figure}

The 2011 outburst of T Pyx provides the community an opportunity for modern,
multi-wavelength and time-series observations of a nova event and will
hopefully constrain the nature of this complex system. Among 

\

\

\

\noindent
these observations, X-ray grating spectra supply information on the highest
ionization states of ejected material and the best constraints on the
post-outburst WD photosphere.

From the day of optical discovery, 2011 April 14.29
\citep{2011IAUC.9205....1W}, T Pyx was monitored with the {\sl Swift}
telescope \citep{2004ApJ...611.1005G}. Figure \ref{fig:slc} displays the {\sl
  Swift} X-Ray Telescope (XRT) count rate (top) and AAVSO V-band magnitude
(bottom) lightcurves. We have plotted the entire 0.3-10 keV range of the {\sl
  Swift} lightcurve (Figure \ref{fig:slc}, top), but the peak flux is
dominated by soft emission in the 0.3-0.8 keV range. A faint X-ray source was
observed at the position of T Pyx \citep{2011ATel.3285....1K} from optical
discovery to 2011 July 31 (day 105) at which point the source was no longer
detectable \citep{2011ATel.3549....1O}. 2011 August 3 (day 111) saw the X-ray
flux begin to rise, peaking on 2011 September 6 (day 145) and plateauing for
around 40 days before declining monotonically. The substantial delay in the
X-ray peak is conventionally attributed to the ejection of optically thick
material in the early phase of the outburst \citep{1994ApJ...437..802K}.  As
the ejected shell expands, it becomes optically thin to X-ray emission
revealing the underlying WD. Because the morphology and time scale of outburst
lightcurves are extremely variable from one RN to the next, the publicly
available {\sl Swift} data provide a fundamental tool for scheduling X-ray
observations.
 
T Pyx had been detected with {\sl Swift} since 2011 August, but due to a
number of different planning and scheduling constraints, X-ray grating
observations were not scheduled till the end of 2011 October, when the average
Swift XRT count rate had decreased from a peak of $\sim$0.6 cts s$^{-1}$ to
$\sim$0.1 cts s$^{-1}$ ({\sl Chandra}, 210 days) and to $\sim$0.02 cts
s$^{-1}$ ({\sl XMM}, 235 days; see Figure \ref{fig:slc}).

Although the XRT plateau suggested a high luminosity, the broad-band data were
not ideal for the identification of the origin of the X-ray emission, whether
super-soft (SSS) X-ray emission from the WD surface or unresolved emission
lines in the ejected shell. This highlights the significance of obtaining
X-ray grating observations during the SSS phase.
\citet{2010ApJ...717..363R}\footnote{http://astro.uni-tuebingen.de/$\sim$rauch/}
calculated a set of publicly available WD atmospheric models that place
constraints on the mass, temperature, and composition of the WD.  These
parameters are extremely valuable in order to evaluate whether RNe are viable
Type Ia supernova progenitor candidates. Having an estimate of the WD mass as
well as spectral features of the circumstellar material are both important to
trace a possible pre-supernova Ia evolution.

In this work we present the two X-ray grating observations, obtained with {\sl
  Chandra} and {\sl XMM-Newton}. We also present the first in a series of
optical spectra taken with the Southern African Large Telescope ({\sl SALT})
during the nova's return to quiescence.  Observations and data reduction
methods are presented in Section \ref{data}. Section \ref{analysis} presents
the analysis of the spectral and temporal components of the X-ray and optical
data. Section \ref{conc} provides a discussion of our results with a
conclusion in Section \ref{conclusion}.

\section{DATA AND REDUCTION METHODS}
\label{data}
In the framework of our pre-approved guest target of opportunity program, T
Pyx was observed with \textit{Chandra} and \textit{XMM-Newton} 210 and 235
days after the 2011 April 14 discovery, respectively. Table \ref{tab1}
provides a summary of the X-ray and optical observations. With each satellite
we utilized dispersion gratings to obtain high spectral resolution
data. Optical spectra of T Pyx were also obtained with the {\sl SALT}
telescope in 2012 November and December.

\begin{figure*}[!th]
  \centering
    \includegraphics[keepaspectratio=true,scale=0.5]
    {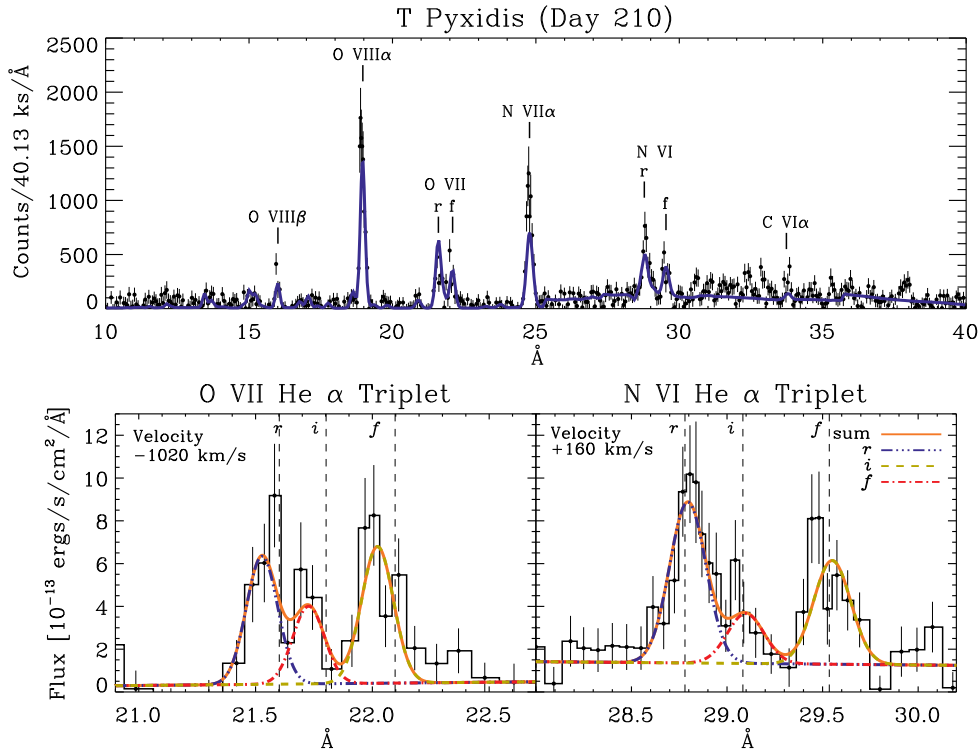}
    \caption{\textit{Top}: Binned \textit{Chandra} spectrum of T Pyx 210 days
      after outburst. XSPEC best-fit model is over plotted in blue (see Table
      \ref{mpar} for model parameters). \textit{Bottom}: Flux corrected O VII
      (left) and N VI triplets (right) with three Gaussian fits. Vertical
      dashed lines represent the rest wavelengths of the $r$, $i$, and $f$
      lines from left to right respectively for each triplet.}
    \label{fig:chandra}
\end{figure*}

\subsection{Chandra}
\label{chandra}
On 2011 November 3 (day 210) we obtained a 40.13 ks \textit{Chandra}
observation of T Pyx using the Low-Energy Transmission Grating (LETG) with the
High Resolution Camera in spectroscopy mode (HRC-S). The +1 and -1 spectral
orders were reduced, extracted, and combined using CIAO (version 4.2).

Even with a modest count rate the spectrum is rich in emission lines and
continuum present from $\sim$25$-$40 \AA \ (see Figure
\ref{fig:chandra}). With a spectral resolution of $R\sim1000$
\citep{1997SPIE.3113..181B}, we are able to resolve a complex line spectrum
with velocity broadening on the order of 2500 km s$^{-1}$ (see Section
\ref{chand}), significantly larger than the instrumental broadening.

The X-ray lightcurve was also extracted from the HRC with CIAO (version 4.2)
and shows a periodic, factor of $\sim$2, variability in the count rate across
the $\sim$40 ks observation (see Section \ref{xlc}).

Near simultaneous {\sl Swift} observations yield an XRT count rate of $0.11\pm
0.01$ and a uvm2 magnitude of $12.23\pm 0.02$. The V-band magnitude at this
time was $11.71\pm0.02$ (AAVSO).

\begin{deluxetable*}{rccccccc}[!th]
\tablewidth{0pt}
\tabletypesize{\small}
\tablecaption{X-Ray Emission Lines}
\tablehead{
  \colhead{Ion} &
  \colhead{$\lambda _{\rm rest}$} &
  \colhead{$\lambda _{\rm obs}$} &
  \colhead{Vel. Centroid} &
  \colhead{FWHM} &
  \colhead{$\lambda _{\rm obs}$} &
  \colhead{Vel. Centroid} &
  \colhead{FWHM} \\
  \colhead{} &
  \colhead{(\AA)} &
  \colhead{(\AA)} &
  \colhead{(km s$^{-1}$)} &
  \colhead{(km s$^{-1}$)} &
  \colhead{(\AA)} &
  \colhead{(km s$^{-1}$)} &
  \colhead{(km s$^{-1}$)} \\
  \hline
  \colhead{} &
  \colhead{} &
  \multicolumn{3}{|c}{Day 210} &
  \multicolumn{3}{|c}{Day 235}
}
\startdata
Ne X Ly $\alpha$  &12.13 & ... & ... &... & ... & ... &...\\
O VIII Ly $\beta$ &16.01 & $15.95\pm0.02$&$-1080\pm320$&$1990\pm710$ & ... & ... &...\\
O VIII Ly $\alpha$&18.97&$18.94\pm0.01$&$-420\pm100$&$2980\pm220$&$19.81\pm0.01$&$-1000\pm90$&$3300\pm200$\\
N VII Ly $\beta$  &20.91 & ... & ... &... & ... & ... &...\\
\\
$r$               &21.60 & $21.53\pm0.01$ & & & ... &  &\\
O VII He $\alpha$ $i$ &21.80 & $21.73\pm0.01$ &$-1020\pm180$& $2190\pm300$ & ... & ... &...\\
$f$               &22.10 & $22.02\pm0.01$ &  &  & ... &  &\\
\\
N VII Ly $\alpha$&24.78&$24.77\pm0.01$&$-190\pm110$&$3010\pm260$&$24.73\pm0.01$&$-580\pm80$&$2850\pm160$\\
\\
$r$               &28.78 & $28.80\pm0.01$ & & & ... &  &\\
N VI He $\alpha$ $i$ &29.08  & $29.10\pm0.01$ & $+160\pm140$ &$2330\pm300$ & ... & ... &...\\
$f$               &29.53 & $29.55\pm0.01$ & & & ... & &\\
\\
C VI Ly $\alpha$ & 33.73 & ... & ... &... & ... & ... &...\\
\enddata
\label{lines} 
\end{deluxetable*}

\subsection{XMM-Newton}
\label{xmm}
The nova was observed again 235 days after outburst with \textit{XMM-Newton}
on 2012 November 28 for 30.92 ks. The observation was conducted with all six
\textit{XMM-Newton} instruments, namely the optical monitor (OM), EPIC-pn, the
two EPIC MOS (all in imaging mode), and the two RGS gratings. The pn was used
in the ``prime small window'' mode; the MOS 1 and 2 were used in ``prime full
window'' and ``prime partial W3'' modes respectively. The thin filter was used
for both MOS and pn.

The data were reduced with the \textit{XMM-Newton} standard analysis system,
XMM-SAS, version 11.0.2. The grating spectra show the same basic spectral
properties as the \textit{Chandra} spectrum of 25 days earlier only with a
decrease in average flux by more than a factor 2. The EPIC-pn camera provided
the highest signal-to-noise ratio (S/N) lightcurve of our observations and
showed the same variability seen in the previous {\sl Chandra} observations
but with lower amplitude (see Section \ref{xlc}). The OM data were taken in
imaging mode with the uvw2 filter with 25 exposures lasting 800-900 s during
the $\sim$30 ks exposure. An average uvw2 magnitude of 12.35 was obtained with
no clear trend or periodicity.

Adjacent {\sl Swift} observations measured uvm2 magnitudes of $12.85\pm0.02$
and $12.71\pm0.02$. The count rate of the nearest XRT observation had dropped
significantly from the {\sl Chandra} observation to $0.024\pm0.005$. The AAVSO
V-band magnitude also fell to $12.06\pm0.01$.

\subsection{SALT}
\label{salt}
The {\sl SALT} telescope located in Sutherland, South Africa with a
$\sim$10.5m diameter mirror is at the moment the largest single disk telescope
\citep{2006SPIE.6267E..32B}. {\sl SALT} spectra, the first in a planned
regular monitoring in the first three years after the outburst, have already
been obtained. Observations utilized the Robert Stobie Spectrograph (RSS)
\citep{2003SPIE.4841.1463B,2003SPIE.4841.1634K} in long slit mode and with the
PG 2300 and PG 900 gratings. Our low-resolution observations were made on 2012
November 22 (day 588) and 2012 December 17 (day 613) covering a 3150-6300 \AA
\ range at a spectral resolution R$\sim$800. A higher resolution spectrum
(R$\sim$3000) in the 4600-5100 \AA \ wavelength range was obtained on 2012
December 6 (day 602).  For information on {\sl SALT}, the RSS, and its
instruments see \citet{2008SPIE.7014E...6B}. The data were reduced and

\

\noindent
analyzed with the {\sl Pyraf} {\sl SALT} reduction pipeline
\citep{2010SPIE.7737E..54C} and {\sl IRAF} \footnote{IRAF is distributed by
  the National Optical Astronomy Observatory, which is operated by the
  Association of Universities for Research in Astronomy (AURA) under
  cooperative agreement with the National Science Foundation.} software
packages.

\section{ANALYLSIS}
\label{analysis}

\subsection{Day 210 X-Ray Spectrum}
\label{chand}
The top panel of Figure \ref{fig:chandra} displays the binned \textit{Chandra}
spectrum 210 days after the outburst discovery with an XSPEC best-fit model in
blue. Expanded regions around the O VII and N VI helium-like triplets are
presented in the bottom panels. The spectrum is dominated by broad emission
lines of H- and He-like ionization states of C, N, and O with low-level
continuum from 25$-$40 \AA.

\subsubsection{XSPEC Spectral Modeling Procedure}
\label{fitp} Emission line and continuum model fitting is done simultaneously
in XSPEC using $\chi^2$ statistics to determine the goodness of fit with the
Churazov weighting scheme \citep{1996ApJ...471..673C}, commonly used in low
count spectral fitting. Models are fit to unbinned data but are presented in a
binned format for clarity. For consistency we also fit the data using the
C-statistic method \citep{1979ApJ...228..939C} at various binnings and find
equivalent results. The final model is presented in as the over-plotted blue
line in the top panel of Figure \ref{fig:chandra}. All components were fit
with a single N(H) value which agrees with reddening estimates and the
expected column density to T Pyx \citep{2013A&A...549A.140S}.

\begin{deluxetable*}{lrrrrrrrrr}[!th]
  \tablewidth{0pt} \tabletypesize{\footnotesize} \tablecaption{Physical
    parameters of the fits shown in Figures \ref{fig:chandra}, \ref{fig:xmm},
    \& \ref{fig:mspec}.}
\tablehead{ 
\colhead{Parameter\,\tablenotemark{a}} &
\colhead{Nov 3 2011} & 
\multicolumn{2}{c}{Nov 28 2011} &\\
\colhead{} &
\colhead{Day 210} &
\multicolumn{2}{c}{Day 235} \\
\colhead{}&
\colhead{}&
\colhead{Fit N(H)}&
\colhead{Fixed N(H)}
}
\startdata
N(H) (10$^{21}$ cm$^{-2}$) & 1.5(0.1) & 1.6(0.1) & 1.5\\
N/H\,\tablenotemark{b} & 15 & 15 & 15 \\
F$_{\rm{Model}}$ (10$^{-12}$ ergs cm$^{-2}$ s$^{-1}$) & 4.08 & 1.81 & 2.04\\

\hline
T$_{\rm WD}$ (K)  & 417,000(22,500) & 405,000(22,500) &  404,000(22,500)\\
R$_{\rm WD}$ $\times$ (d/4.8kpc) (km) & 3,050(1,670) & 3,050 & 3,050 \\
F$_{\rm WD}$ (10$^{-12}$ ergs cm$^{-2}$ s$^{-1}$)\,\tablenotemark{c} & 2.15 & 0.61 & 0.90\\
F$_{\rm WD,{\it un}}$ (10$^{-12}$ erg cm$^{-2}$ s$^{-1}$)\,\tablenotemark{d} & 53.8 & 40.61 & 40.34\\

\hline
kT$_{\rm P,1}$ (eV) & 94(13) & 116(3) & 115(3)\\
Vol. Em. Measure $\times$ (d/4.8kpc)$^2$ (10$^{55}$ cm$^{-3}$)&51(43)&36(6)&30(4)\\
v (FWHM, km s$^{-1}$) & 2,380(390) & 2,360(250)& 2,350(250)\\
Centroid Shift (km s$^{-1}$) & 90(570) & -30(1,360) & -310(1510)\\

\hline
kT$_{\rm P,2}$ (eV) & 240(10) & 250(10) & 250(10)\\
Vol. Em. Measure $\times$ (d/4.8kpc)$^2$ (10$^{55}$ cm$^{-3}$)&28(8)&14(1)&12(1) \\
v (FWHM, km s$^{-1}$) & 2,870(180) & 2,690(190) & 2,680(190)\\
Centroid Shift (km s$^{-1}$) & -150(2,580) & -610(1,870) & -990(2,080) \\

\hline
kT$_{\rm P,3}$ (eV) &  & 1,260(30) & 1,250(40) \\
Vol. Em. Measure $\times$ (d/4.8kpc)$^2$ (10$^{55}$ cm$^{-3}$)& &6(1)&5(1)\\
\enddata

\tablenotetext{a}{Component order: Tuebingen-Boulder ISM Absorption Model,
  Rauch WD atmosphere, BVAPEC 1, BVAPEC 2, BVAPEC 3. 1$\sigma$ confidence
  levels are given in parenthesis. } 
\tablenotetext{b}{N/H indicates nitrogen mass fraction compared to solar.}
\tablenotetext{c}{Fluxes are measured in the 0.25-1.24 keV range.} 
\tablenotetext{d}{The ``{\it un}'' subscript refers to the ``unabsorbed'' flux.}  
\label{mpar} 
\end{deluxetable*}

\subsubsection{Emission Lines}
\label{elines}
Table \ref{lines} presents the line center, velocity centroid, and full width
at half maximum (FWHM) for the most prominent X-ray emission lines. Detected
emission lines with lower S/N are only listed as being present but no attempt
was made to characterize them. These estimates were made by fitting a Gaussian
profile to each flux corrected line profile.  Fits were made to data binned to
a minimum of 20 counts per bin to justify $\chi^2$ minimization and increase
S/N. As a result, the bin size presented in Figure \ref{fig:chandra} (bottom)
is variable with a mean width of $\sim$0.1 \AA. Flux conversion from
instrumental counts was done using the ISIS software package {\sl
  flux\_corr}. This preforms a proper division of data by the auxiliary
response file while accounting for the line shape by also integrating over the
redistribution matrix file. ISIS flux conversion can have an appreciable
uncertainty in absolute flux over a broad wavelength range, however, the
narrow He-like triplet ranges ($\sim$2 \AA) over which we are concerned with
relative flux ratios, are robust (see below). We note that while the intrinsic
line shape may not be Gaussian in nature, at the resolution of our data it
provides a good first order representation.

The He-like triplets of oxygen and nitrogen show blending between the
resonance ($r$), intercombination ($i$), and forbidden ($f$) lines (see Figure
\ref{fig:chandra} bottom panel, labeled vertical lines). We fit the blends
simultaneously with three Gaussians requiring a constant FWHM and velocity
shift across the triplet. We justify this constraint as it is expected the
emitting material is under similar physical conditions. This provides
reasonable agreement with centroid shifts and line widths of unblended H-like
lines. Data for He-like triplets in Table \ref{lines} are presented as the
best fit velocity centroid and FWHM of the three lines.

Blends from other emission lines may be present in the data contaminating our
estimates from Gaussian fitting such as, Fe XVIII (16.20 \AA) with O VIII Ly
$\beta$ (16.01) and the N VI He $\beta$ resonance line ($\sim$24.89 \AA) with
N VII Ly $\alpha$. The former we conclude to be negligible due to the absence
of other Fe emission lines in the spectrum. The latter is likely to have some
level of contamination but given the plasma temperature we derive from the N
VII He-like triplet (see below), the expected contribution of N VI He $\beta$
(from an APEC model) is meager compared to N VII Ly $\alpha$.

In order to describe the ionization and excitation mechanism behind the
observed emission lines, as well as placing estimates on the electron
temperature and density of the emitting material, we analyze the He-like
triplet flux ratios of N VI and O VII using diagnostics from
\citet{2001A&A...376.1113P} and \citet{1972ApJ...172..205B}. We calculate the
$G$ ratio, $(f+i)/r$, which is sensitive to the ionization mechanism and
electron temperature, where $G$ ratios $\sim$1 indicate a plasma in
collisional ionization equilibrium (CIE). Larger values indicate contributions
of photoionization \citep{2000A&AS..143..495P}. We also utilize the $R$ ratio,
$f/i$, which is sensitive to electron density. Triplet diagnostics require
energy flux ratios, not instrumental counts, hence the need for ISIS flux
conversion. From the N VI triplet, we calculate a $G$ ratio of
$G_{N}=1.0\pm0.2$ which is consistent with a plasma in CIE at
$T_e=1.1^{+0.2}_{-0.2}\times10^6\, $K.

The $R$ ratio yields $R_{N}=2.1 \pm0.8$, which corresponds to a density of
$n_e=8\pm6\times10^{9}$ cm$^{-3}$ assuming only collisional excitation. The
$R$ ($f/i$) diagnostic can be strongly influenced by UV radiation, which
excites $f$ level electrons into the $i$ level decreasing the $R$ ratio and
suggesting erroneously large densities. This effect is extreme in cases where
the emitting material is near a strong UV emitter (e.g., O star winds where
$R$ is typically $\sim$1). In the case of T Pyx, we conclude that our result
is largely unaffected by UV radiation, given that: (1) the UV flux measured
with {\sl Swift} is low at this stage of the outburst, (2) the emitting
material is in the ejecta and likely far from the WD (at a maximum outflow
velocity of $\sim$4000 km s$^{-1}$ found by \citet{2011A&A...533L...8S}, a
distance $>10^{10}$ km is found for a constant velocity outflow
\citep{2013A&A...549A.140S} at day 210), and (3) the $R$ value obtained falls
in the steepest portion of the $R({\rm n}_e)$ curve where electron density is
relatively insensitive to $R$ (see \citet{2000A&AS..143..495P}; Figure
8). With only loose constraints on the UV flux and the distance of the
emitting material, this density should be considered an upper limit. Results
are, however, consistent with values derived for other novae (e.g., U Sco,
\citealt{2013MNRAS.429.1342O}; Nova Cygni 1975,
\citealt{1978ApJS...38...89N}).

\

Applying the same method to the O VII triplet, we find a $G$ ratio of
$G_{o}=1.7 \pm0.5$, and an $R$ ratio of $R_{o}= 1.8 \pm0.7$. These estimates
provide an electron temperature of $T_e=0.5^{+0.5}_{-0.2}\times10^6$ K and a
density of $n_e=3\pm3\times10^{10}$ cm$^{-3}$, again, assuming the plasma is
in CIE with no contamination of UV flux. While these values agree with those
from the N VI triplet, we note the lower temperature derived from O VII. In
contrast, the O VIII Ly $\beta$ (16.01 \AA) to Ly $\alpha$ (18.97 \AA) ratio
gives a temperature of $T_e>2.5\times10^{6}\, $K (assuming CIE;
\citealt{2001ApJ...556L..91S}).

\begin{figure*}[!th]
  \centering
    \includegraphics[keepaspectratio=true,scale=0.5]
    {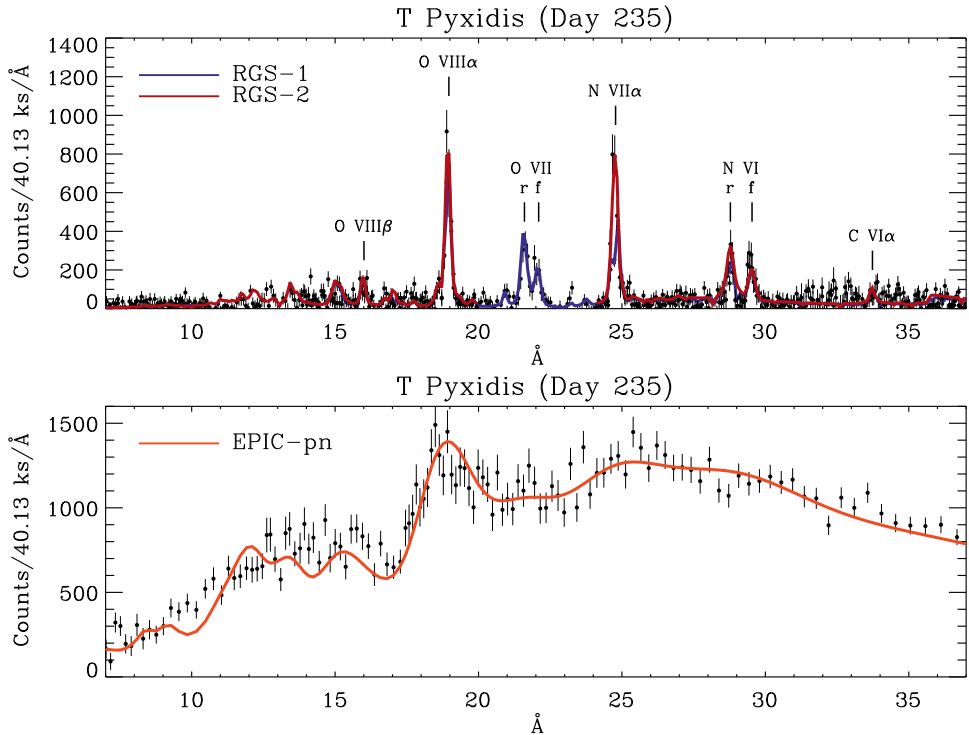}
    \caption{Binned {\sl XMM-Newton} spectra 235 days after
      discovery. \emph{Top}: RGS-1 and RGS-2 spectra with model overlaid in
      blue and red respectively (model ``Fit N(H)'' from Table
      \ref{mpar}). \emph{Bottom}: Concurrent EPIC-pn spectrum with final model
      over-plotted. Presented model was fit simultaneously to RGS 1 and 2 and
      EPIC-pn data.}
    \label{fig:xmm}
\end{figure*}

The discrepancy of the O VIII Ly$\beta$/$\alpha$ and O VII triplet electron
temperature diagnostics, may be due to many factors that can influence the
relative flux ratios of He-like triplets. Firstly, we note that the O VII $G$
ratio is elevated for a purely collisionally ionized plasma (usually
$\sim$1). One possibility, if the ratio is unaltered (by scattering for
instance), is the presence of partial photoionization. This seems unlikely,
however, given the minimal continuum flux in the region. Alternatively,
geometry dependent resonance scattering can also effectively increase or
reduce the $r$ line affecting the $G$ ratio
\citep{2010SSRv..157..103P}. Modeling the scattering plasma geometry is beyond
the scope of this paper and neither effect can be definitively ruled out.

Given the above discussion, we adopt the O VIII Ly $\beta/\alpha$ ratio and N
VI triplet temperature diagnostics assuming a plasma in CIE with multiple
temperature components. The N VI triplet diagnostic is more robust than O VII
because, even near the continuum emission, it shows no signature of
photoionization and if this system is at all similar to the 2010 outburst of
RN U Sco, softer lines are emitted further from the WD than harder lines,
decreasing the possibility of scattering
\citep[see][]{2012ApJ...745...43N}. WD continuum models from
\citet{2010ApJ...717..363R} also predict negligible atmospheric absorption at
these continuum levels compared to the observed line flux.  Requiring multiple
plasma temperature components is not unexpected given the spatial complexity
of the ejecta \citep{2013A&A...549A.140S} and broad range of ionization
potentials observed.

Informed by our flux corrected triplet diagnostics, we apply an APEC model
\citep{2001ApJ...556L..91S} in XSPEC at the N VI temperature ($10^{6}\, $K)
for the initial model. The APEC model assumes optically thin plasma in CIE.
To account for the non-solar abundances of novae ejecta and in order to fit
the velocity broadening of the lines, we chose the APEC model that includes
velocity broadening and ad-hoc abundances of different elements (BVAPEC).  We
find a good fit to the softer lines redward of 25 \AA \ with an increased
nitrogen abundance ($\sim$15 times the solar N/H mass ratio). A high N
abundance is expected for CNO-cycle processed material so this rough estimate
seems a reasonable value for modeling purposes.

The best fit model has a velocity broadened FWHM of 2400$\pm200$ km s$^{-1}$
(similar to the low energy lines of Table \ref{lines}) and a centroid shift
consistent with zero. This single model component however, greatly
underestimates the emission lines blue-ward of 25 \AA \ and thus confirms the
need for two plasma components.
 
As a second component in the fit, we add an additional variable abundance,
velocity broadened, BVAPEC model with the O VIII Ly $\beta/\alpha$ plasma
temperature ($2.5\times10^{6}\, $K) and find a good fit to the harder emission
lines with the same enhanced nitrogen abundance and velocity broadening of
2600$\pm200$ km s$^{-1}$. This additional model component reduced $\chi^2$
from 1.10 to 1.07 (including a continuum model, see below).

\

The blueshift and broadening parameters of these model components are less
than, but in general agreement with the trends of Table \ref{lines}. The
uncertainties on the velocity widths and centroid shifts are large owing to
the low S/N of the data and the attempt to fit a single set of parameters to
lines of varying widths and centroid shifts (see Table \ref{mpar}).

\begin{figure*}[!th]
  \centering
    \includegraphics[keepaspectratio=true,scale=0.5]
    {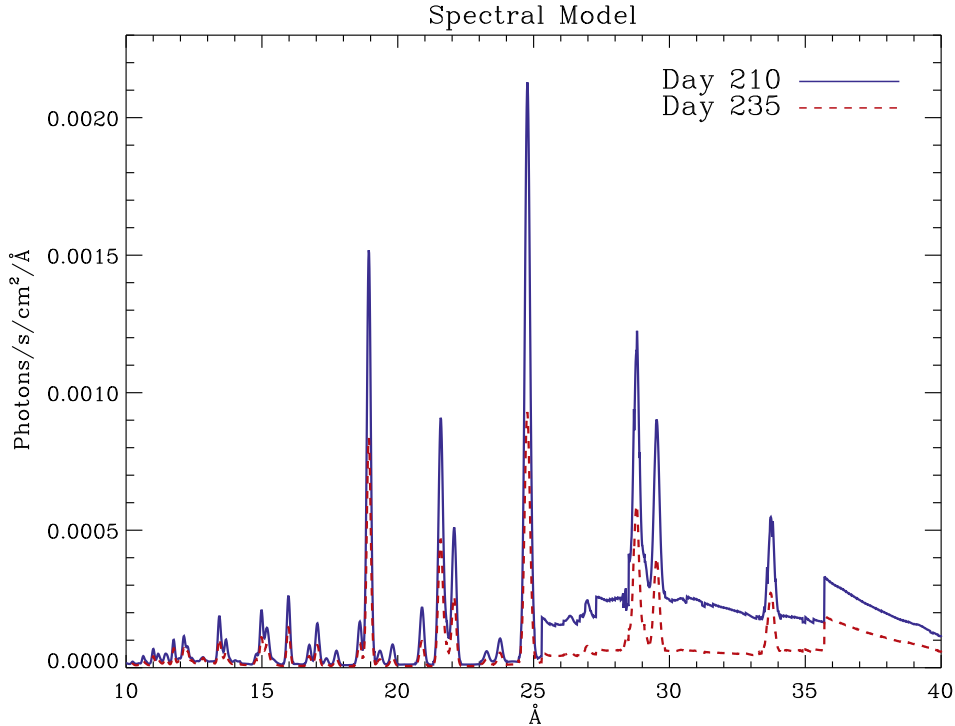}
    \caption{Evolution of the XSPEC X-ray best-fit model from 210 (blue solid)
      to 235 (red dotted) days after discovery. See Table \ref{mpar} for
      best-fit parameters, day 235 is the middle, ``Fit N(H),'' column.}
    \label{fig:mspec}
\end{figure*}

\subsubsection{Continuum}
\label{ccont}
From the BVAPEC models above, the observed continuum emission is too large to
be due to Bremsstrahlung alone. Although an ad-hoc Bremsstrahlung component
can also produce a higher continuum in the right range, it does not yield a
good fit.

To describe the continuum we used WD atmospheric models
\citep{2010ApJ...717..363R} and found good fits to the data with reduced
$\chi^2\simeq 1$. We fitted models using grids of log($g$), with the given
steps of 0.5 in log($g$). We found that models with log($g$)$<$8 do not fit
the data. Models with log($g$)$\ge$9 reach the lower bound (500,000 K) of the
physical temperature range before converging on a significant fit. Since the
\citet{2010ApJ...717..363R} models are most discrepant in the presence of
strong absorption features and flux below 20 \AA, of which we observe neither,
we obtain almost equally good fits with log($g$)$=$8 and 8.5. Given that there
is no significant difference in the derived temperature between these models,
we present here results with log($g$)=8, closer to values indicated in the
theoretical outburst model for T Pyx \citep{2012BaltA..21...76S}.

The atmospheric models with log($g$)=8 yield a best fit with T$_{\rm
  eff}\sim420,000$ K, and a N(H) consistent with reddening estimates from
\citet{2013A&A...549A.140S}. Emission from this continuum model sharply cuts
off at wavelengths short of $\sim 25$ \AA. WD models of lower temperature
quickly begin to underrepresent the continuum flux in the 25-30 \AA \
range. This final model with the emission line components detailed above is
presented in blue in the top panel of Figure \ref{fig:chandra}.  Table
\ref{mpar} contains full model parameters with 1$\sigma$ confidence levels in
parenthesis. In all parameters we have chosen a distance to T Pyx of 4.8 kpc
based on recent light echo observations with the {\sl Hubble Space Telescope}
Wide Field Camera 3 \citep{2013ApJ...770L..33S}.

We obtain an unabsorbed X-ray luminosity of 1.48 $\times 10^{35}$ erg s$^{-1}$
in the 0.25-1.24 keV range. This corresponds to a bolometric luminosity of
only $\sim$10$^{36}$ erg s$^{-1}$ while the expected luminosity for a WD at
this temperature is L$_{Bol}$$\geq 3\times10^{37}$ erg s$^{-1}$ corresponding
to a mass $\leq$ $\sim 1$M$_\odot$ \citep{2012BaltA..21...76S}. Given the lack
of detected absorption features in our spectra we can provide only an upper
limit on the WD effective temperature while log($g$) and surface abundances
remain only loosely constrained.

Because T Pyx is thought to be a low inclination system, an explanation of the
low luminosity in terms of a hidden WD seen through Thomson scattering corona,
like in U Sco \citep{2012ApJ...745...43N,2013MNRAS.429.1342O}, does not
apply. The flux we measure would be consistent only with a partially obscured
WD with ejecta that is still partly optically thick to the X-rays. We discuss
this explanation more in Section \ref{conc}.

\subsection{Day 235 X-Ray Spectra}
\label{xmmsec}
Although the average flux is lower by more than a factor of two in the {\sl
  XMM-Newton} spectra, it is reduced by as much as 70\% in the 30-40 \AA \
range, where the WD continuum dominates. Figure 3 displays the two RGS grating
spectra on top and the EPIC-pn spectrum on bottom with best fit models
overlaid. Interestingly, the peak in the continuum and the flux from 25$-$30
\AA \ is still present with the same spectral shape only at decreased
levels. Qualitatively, this would require a decrease in flux while maintaining
the WD temperature.

We simultaneously fit the RGS-1, RGS-2, and EPIC-pn spectra, shown in Figure
\ref{fig:xmm}, with the method described in Section \ref{fitp}. Implementing
the best fit parameters from day 210 as an initial model we find that the
continuum behavior can be reproduced in three ways that are degenerate to our
data. First, a 10\% increase in the model N(H) with a 3\% drop in WD
temperature (within 1$\sigma$ confidence limit of day 210 fit temperature) and
constant WD radius (model ``Fit N(H)'' in Table \ref{mpar}). Second, a
slightly larger decrease in WD temperature (again smaller than 1$\sigma$
confidence level) while N(H) and radius remain constant (model ``Fixed N(H)''
in Table \ref{mpar}). Third, a $\sim$20\% decrease in the WD radius while WD
temperature and N(H) remain constant (not included, see Section
\ref{conc}). Insignificant statistical difference is found between these
models and all produce acceptable fits. Figure \ref{fig:mspec} shows the two
best-fit models for day 210 and day 235 (``Fit N(H)'' in Table
\ref{mpar}). The spectral shape between models varies little and the ``Fit
N(H)'' model from Table \ref{mpar} is presented in Figures \ref{fig:xmm} and
\ref{fig:mspec}.

\begin{figure*}[!tbh]
  \centering
    \includegraphics[keepaspectratio=true,scale=0.60,angle=90.0]
    {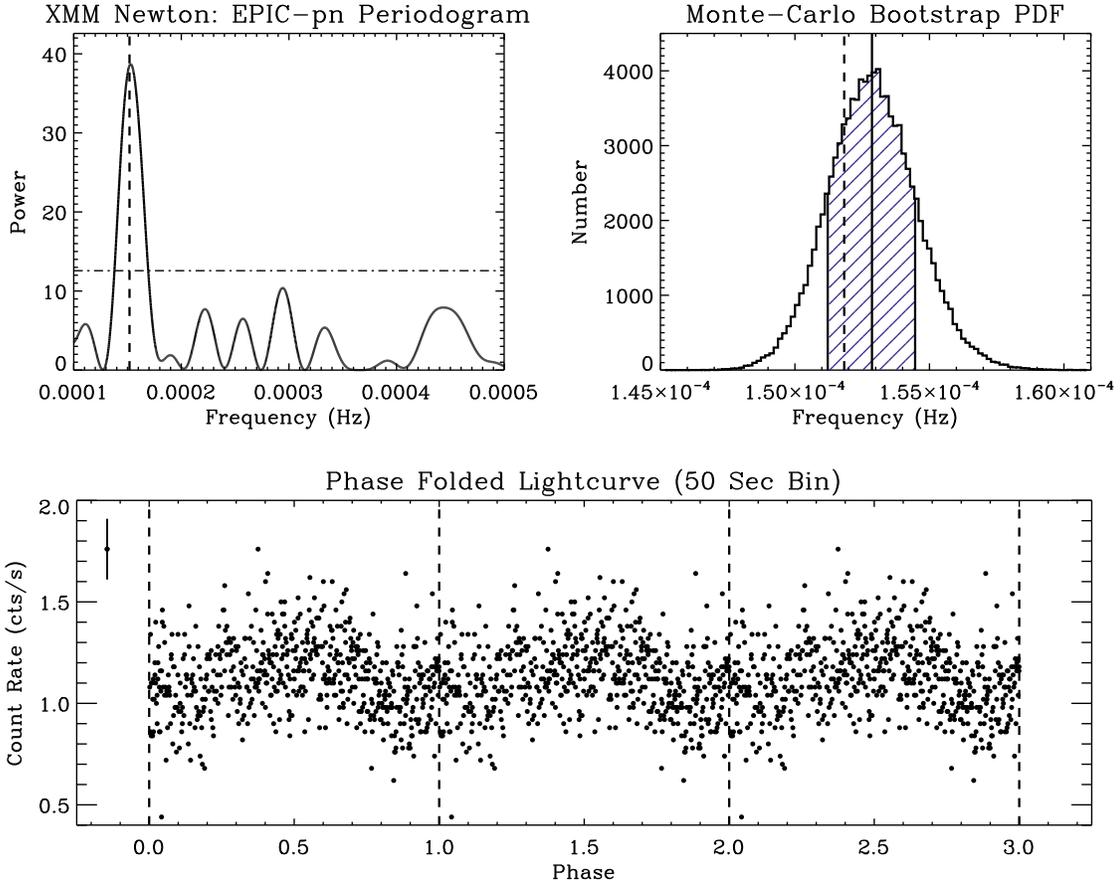}
    \caption{\textit{Top Left}: Lomb-Scargle periodogram of the
      \textit{XMM-Newton}, EPIC-pn lightcurve (235 days after
      outburst). Dot-dashed horizontal line displays the 1\% False Alarm
      Probability level. Dashed vertical line represents the extrapolated
      optical period for the epoch of observation (1.830 hr). \textit{Top
        right}: Probability Distribution Function of Monte-Carlo bootstrap
      error simulation. Solid vertical line represents periodogram peak from
      left panel (1.817$\pm$0.019 hr). Dashed vertical line again represents
      expected optical period. The shaded region encloses 68.3\% of the error
      distribution. \textit{Bottom}: Phase folded lightcurve of data binned at
      50 s. Leftmost point displays the average error bar, phase zero is set
      to the optical ephemeris time. Data are repeated over three phases to
      highlight modulation pattern.}
    \label{fig:pnlc}
\end{figure*}

The two plasma components show little variation in temperature but a decrease
in flux from 25 days earlier. Parameters for centroid shift and velocity
broadening remain uncertain as in the {\sl Chandra} spectrum. Line ratios and
the $G$ and $R$ quantities mentioned above appear roughly the same but as 25
days earlier but due to the reduced S/N, no attempt was made to fit them. Fits
to 

\

\

\noindent
the N VII Ly $\alpha$ and O VIII Ly $\alpha$ lines are presented in the
three right columns of Table \ref{lines}).

The inclusion of the EPIC-pn data revealed an extra source of hard emission in
the 5-15 \AA \ range that was under represented by the initial model from day
210. We find that this emission can be modeled with a third plasma component
at $T_{\rm e}=\sim15\times10^{6}$ K with similar plasma characteristics as the
lower temperatures components. The values of this model component are less
certain because it is mainly constrained by the EPIC-pn broadband data. The
addition of this component improved the reduced $\chi^2$ by 0.13 to 0.99. We
do not believe this marks the emergence of a harder emission component, only
that it went undetected in the Chandra (day 210) spectrum due to low S/N.

\begin{figure*}[!ht]
  \centering
  \includegraphics[keepaspectratio=true,scale=0.60]{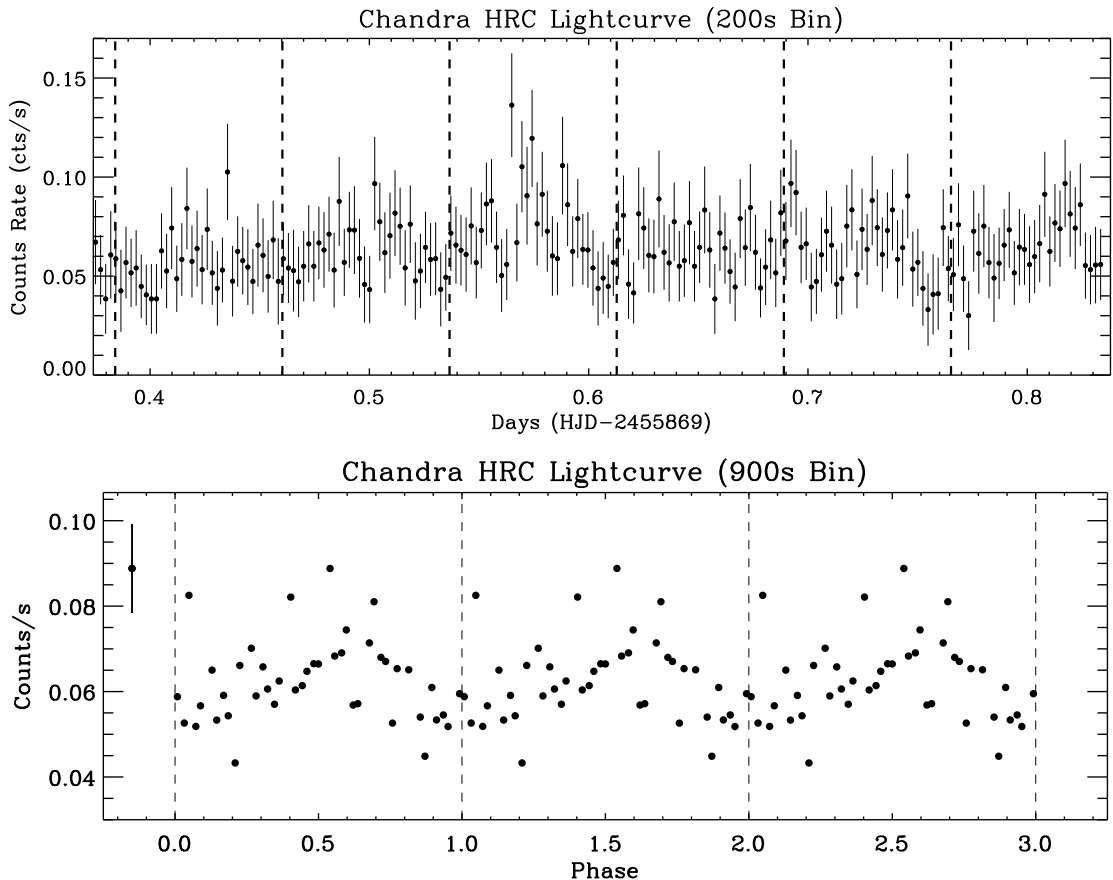}
  \caption{\textit{Top}: {\sl Chandra}-HRC lightcurve 210 days after
    outburst. Vertical dashed lines are predicted optical ephemerides from
    Equation \ref{eqn:eph}. \textit{Bottom}: Phase folded light curve of
    independently found period (agrees with extrapolated optical
    period). Phase zero is set to the optical ephemeris time.}
  \label{fig:hrc}
\end{figure*}

\

\subsection{X-Ray Lightcurve and Periodicity}
\label{xlc}
The long seen periodic variability in non-outburst optical photometry of T Pyx
was determined to be orbital in nature by \citet{2010MNRAS.409..237U}. These
authors applied the most recent optical photometric ephemeris timing from
ongoing work by \citep{2013arXiv1303.0736P}, updated from

\

\

\noindent
\citet{1998PASP..110..380P}, with the help of additional observations from the
Center for Backyard Astronomy. Timing for the optical ephemeris has the form,
\begin{multline}
\mbox{Minimum Light (HJD)} = 2451651.65255(35)+\\
0.076227249(16) \times N+2.546(54) \times 10^{-11} N^{2},
\label{eqn:eph}
\end{multline}
where $N$ is the number of periods since the given epoch. Numbers in
parenthesis are the error of the last two significant digits
\citep{2010MNRAS.409..237U}. The $N^2$ term is for a lengthening period,
believed to be due to a high mass accretion rate \citep{1998PASP..110..380P}
and was required for agreement with the spectroscopic period.

We compute a Lomb-Scargle periodogram \citep{1982ApJ...263..835S} of the
heliocentric corrected EPIC-pn, MOS, and HRC-LETG lightcurves. The top left
panel of Figure \ref{fig:pnlc} displays the EPIC-pn periodogram. The main peak
is well above the $1\%$ false-alarm-probability (FAP, horizontal dashed
line). The vertical dashed line represents the extrapolated optical frequency
for our epoch of observation from Equation \ref{eqn:eph}. While only a small
frequency window is displayed, no other frequencies produced statistically
significant power.

\begin{figure}[tbh]
  \centering
    \includegraphics[keepaspectratio=true,scale=0.38,angle=90]
    {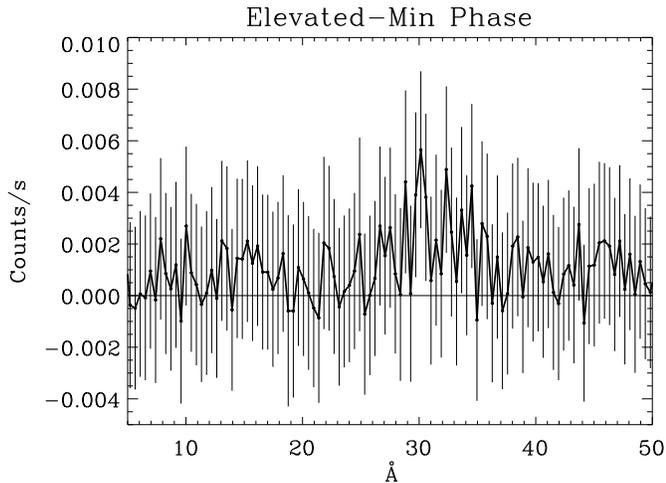}
    \caption{Difference of {\sl Chandra} spectrum extracted in high and low
      phases. Elevated spectra includes phase 0.4-0.7, minimum spectra
      includes phases 0.0-0.15; 0.85-1.0 (compare phase intervals with Figures
      \ref{fig:pnlc} and \ref{fig:hrc}). Result is binned by a factor of 30
      from native resolution.}
    \label{fig:phasecomp}
\end{figure}

To estimate the error in our peak frequency, we perform a Monte-Carlo
bootstrap error simulation of 100,000 iterations. The resultant probability
distribution function is presented in the top right panel of Figure
\ref{fig:pnlc}. The solid vertical line displays the periodogram peak (from
the left panel; Figure \ref{fig:pnlc}) and the dashed line, again, represents
the extrapolated frequency from Equation \ref{eqn:eph}. The hatched region
marks the $68.3\%$ (1$\sigma$) enclosure of the cumulative distribution
function (not shown). Our result is an X-ray period of $1.817\pm0.019$ hr
($0.07571\pm0.00081$ days), which is consistent with the predicted optical
period for this epoch, 1.830 hr. While we only present the full EPIC-pn timing
analysis herein for brevity, the same analysis was carried out for the MOS and
HRC-LETG lightcurves. Both independently recovered the optical period of the
above analysis but with larger uncertainties owning to their lower
sensitivities. We note that prior to the outburst \citet{2012arXiv1208.6120B}
recovered the optical period in X-rays during quiescence using the {\sl
  Chandra} ACIS-S detector.

Recovering the same period at X-ray and optical wavelengths places constraints
on the physical process behind the variability. The bottom panel of Figure
\ref{fig:pnlc} presents the EPIC-pn phase folded lightcurve (day 235) in 50 s
bins. The left-most point displays the average error bar. The optical
ephemeris time is set to phase zero (plotted as vertical dashed lines) where
there is a clear match to the X-ray minimum.

In the top panel of Figure \ref{fig:hrc} the {\sl Chandra} HRC (day 210)
lightcurve is presented with the optical ephemerides over-plotted in dashed
vertical lines. The bottom panel displays the phase folded lightcurve about
its independently measured, consistent period. The noise is much higher
compared to the EPIC-pn but shows the same match to the optical ephemeris. We
note that individual 

\begin{figure*}[!tbh]
  \centering
    \includegraphics[keepaspectratio=true,scale=0.55]
    {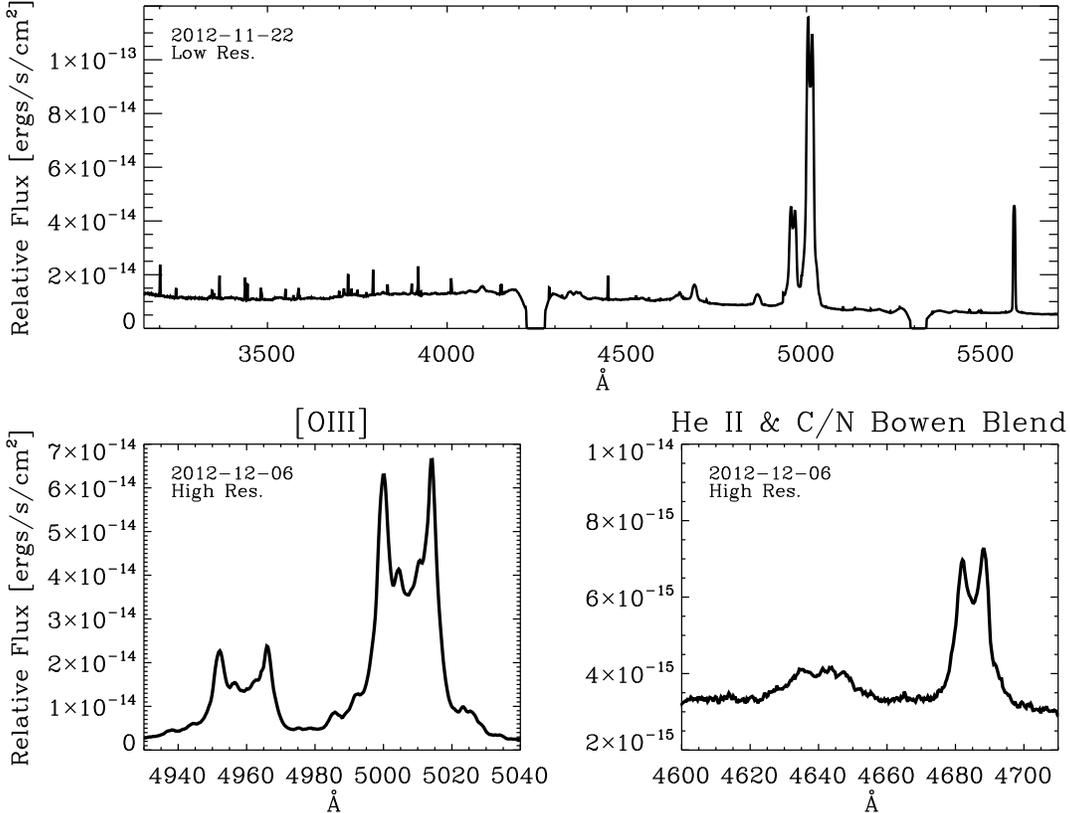}
    \caption{{\sl SALT} optical spectra of the T Pyx nebular
      phase. \textit{Top}: Full low-resolution spectrum from 2012 November
      22. \textit{Bottom}: Regions of high-resolution spectrum from 2012
      December 6. [O III] lines 4960 \AA \ and 5007 \AA \ are presented in the
      left panel with He II 4686 \AA \ and C/N Bowen blend (4640-4650 \AA) in
      the right. Flux is arbitrarily normalized but consistent across bottom
      panels. Wavelength scales are equivalent in bottom panels but notice
      difference in flux scale and profile shape.}
    \label{fig:salt}
\end{figure*}

\

\noindent
periods show significant variability which can be seen in
the sharp rise at $\sim$0.57 and $\sim$0.69 HJD(-2,455,869) of the HRC
lightcurve.

To investigate the spectral region of variability, we extract the {\sl
  Chandra} spectrum (highest S/N) at elevated (0.4-0.7) and minimum
(0.85-1.15) phases and present the differenced result in Figure
\ref{fig:phasecomp}. The variability occurs predominantly red-ward of the N VI
triplet emission lines and is primarily in the continuum flux. In Section
\ref{conc} we comment on possible sources of the periodic variation
originating either from occultation on the suggested orbital period or instead
due to the WD spin.

\subsection{SALT Optical Spectra}
\label{saltsec} 

Two groups found that the mass ejected in the 2011 T Pyx outburst was
``bipolar'' in nature, \citet{2011A&A...534L..11C} only a few days and six
weeks after the outburst, and \citet{2013A&A...549A.140S} six months and one
year post-outburst. This structure, if it persisted in November of 2011, may
well explain a partial obscuration of the WD surface inferred from our X-ray
grating spectra. In order to assess whether the ejecta expansion continued
without spherical symmetry and to obtain further insight into the geometry and
evolution of the shell, we obtained additional optical spectra at a later
epoch. We observed the nova with {\sl SALT} and the PG 2300 grating
(R$\sim$3000) in the 4480-5516 \AA \ region on 2012 December 6, and at lower
resolution with the PG 900 grating (R$\sim$800) in the 3156-6318 \AA \ region
on 2012 November 22 and on 2012 December 17.  There was no significant change
in the spectra between these dates.

\begin{figure*}[!tbh]
  \centering
    \includegraphics[keepaspectratio=true,scale=0.31]{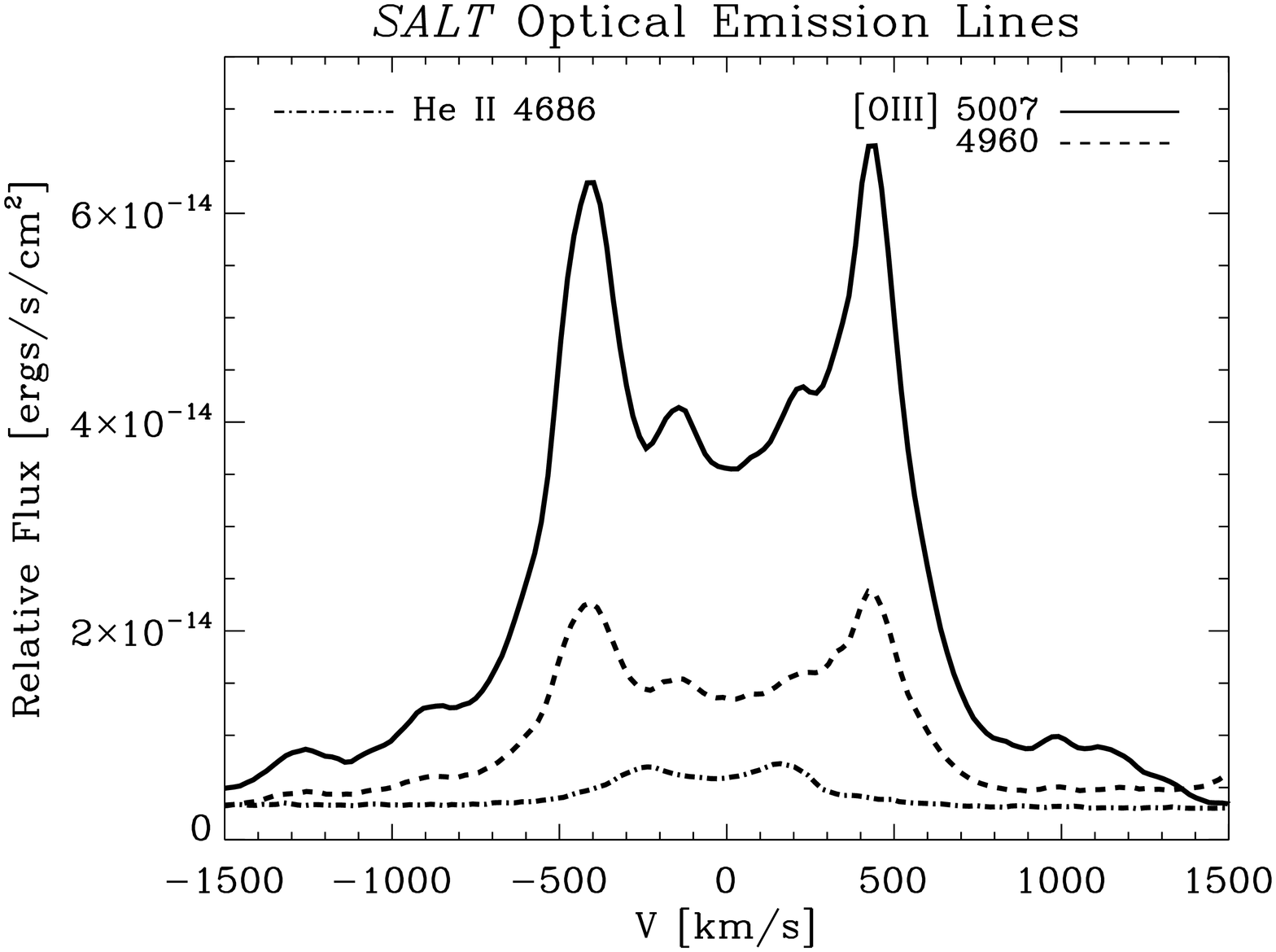}
    \includegraphics[keepaspectratio=true,scale=0.31]{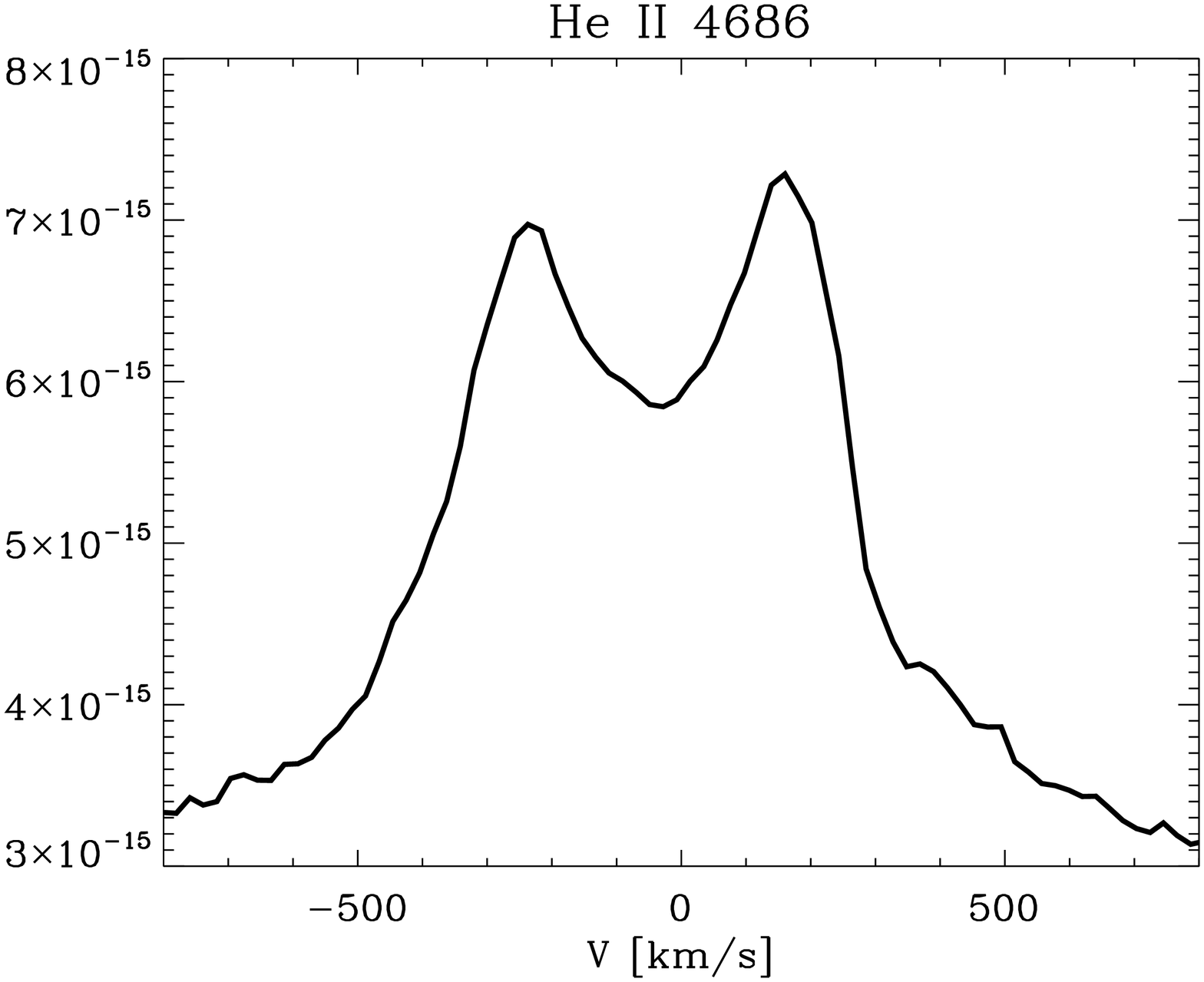}
    \caption{\textit{Left}: {\sl SALT} velocity spectrum of [O III] 5007 \AA \
      and 4960 \AA \ and He II 4686 \AA \ from the high resolution spectrum of
      2012 December 6. \textit{Right}: Expanded velocity region around He II
      4686 \AA \ highlighting the difference in velocity structure of He II to
      [O III].}
    \label{fig:saltv}
\end{figure*}

We found that in December of 2012 (602 days after outburst), T Pyx was still
in the {\sl nebular phase} due to the strong emission of nebular forbidden
lines of [O III] \citep{1990LNP...369..188S}. Figure \ref{fig:salt} shows a
full, low-resolution spectrum from 2012 November 22 in the top panel.  An
enlarged region of the higher resolution spectrum from 2012 December 6 around
the [O III] doublet (4959, 5007 \AA) and He II (4686 \AA) and C/N Bowen blend
are presented on the bottom left and right respectively. The C/N Bowen blend
(4640-4650 \AA) consists of several lines of N III and C III which are seen in
cataclysmic variables and low-mass X-ray binaries with high mass accretion
rates and in novae during outburst. Relative flux calibration across the
spectrum is presented in the bottom panel of Figure \ref{fig:salt} but the
scale is arbitrary.

We compare the equivalent width ratio of the above lines to H$\beta$ with
other recent novae (V382 Vel, V4743 Sgr, and KT Eri) also observed with {\sl
  SALT} ranging from 2.5 to 15 yr after outburst (M. Orio et al. 2013, in
preparation). We find that the [O III] line intensity in T Pyx is still very
large and does not decay as expected \citep{2001JAD.....7....6D}. This is
interesting considering that in most RN, ``[O III] 5007 \AA \ is usually
inconspicuous or absent'' \citep{2001JAD.....7....6D} and generally only in
``very slow novae'' is the [O III] emission prominent more than a year after
the outburst. The long nebular phase is thus unexpected in T Pyx not only
because it is not as slow as most novae with long lasting nebular lines, but
also because of the small amount of ejecta mass that has been estimated
(e.g. Shore et al. 2013). Novae with comparable velocity and ejecta mass
appear to have a much shorter nebular phases. This will be discussed more in
detail in an upcoming article (M. Orio et al. 2013, in preparation).

The ratio of the equivalent width of the He II line at 4686 \AA \ to that of H
$\beta$ is lower than it has been observed in novae that have been super-soft
X-ray sources with effective temperatures above 500,000 K.  Moreover, in the
sample of objects studied by M. Orio et al. (2013, in preparation), the He II
line at 4686 \AA \ is still more prominent than H$\beta$ even 14 yr after the
outburst. The ratio of these lines depends mainly on the temperature of the
ionizing source and the amount of total ionizing radiation absorbed by the
nebular material \citep{1987A&A...181..365S}. Assuming that these lines trace
the same ejected material, this measurement may imply that the super-soft
X-ray source was less hot and potentially cooled more rapidly in T Pyx than in
other recent novae observed as super-soft X-ray sources.

Unlike in other RN at this post-outburst epoch, the [O III] line at 5007 \AA \
has more than 10-fold the flux of the He II line
\citep{2001JAD.....7....6D}. The [O III] lines are due to forbidden
transitions which require a critical density of n$_{\rm e}\leq 10^5$
cm$^{-3}$. The nebular material is far from being spherically distributed and
has a low filling factor $f$, which diagnostics from
\citet{2011A&A...533L...8S} estimate to be $f$=0.03.

The spectral resolution of the PG 900 grating is only of about 6.5 \AA \ at
the wavelength of the [O III] lines, so we will focus on the structure of the
lines observed with the PG 2300 grating, which has a resolution of 1.5 \AA \
(Figure \ref{fig:salt}, bottom). Figure \ref{fig:saltv} displays the velocity
structure of the [O III] and He II lines in the left panel with an enlarged
region around the He II line in the right panel.  In our spectra, the line
profiles still have nearly the same structure detected in 2011 October (day
180) and modeled as bipolar by \citet{2013A&A...549A.140S} with an inner
structure apparently due to a separate bipolar ring at a different opening
angle. Both observations have similar relative weights between the broader and
narrower bipolar components forming the lines at this late stage (see Figure
11 of \citealt{2013A&A...549A.140S}). The blue wing of the line is more
prominent than the red wing, possibly indicating a departure from the
axisymmetric conical geometry.

The He II line at 4686 \AA \ is much less prominent and has a markedly
different velocity structure than the [O III] lines. A double Gaussian fit to
He II 4686 \AA \ yields a FWHM of 328 and 317 km s$^{-1}$ and separation of
384 km s$^{-1}$. These values are comparable to the average FWHM of 450 km
s$^{-1}$ and separation of 300 km s$^{-1}$ used by \citet{2010MNRAS.409..237U}
to describe pre-outburst He II disk emission. It is possible this emission is
coming from a reforming accretion disk. Since the spectra we observed with
{\sl SALT} are very different from the pre-outburst spectra, we assume that
the disk was disrupted or at least disturbed in the outburst. The C/N Bowen
blend in particular is much less prominent in our spectra, while He I emission
features at 4921 \AA \ and 5015 \AA \ are absent. These lines were dominant in
the pre-outburst spectra, and were attributed to the accretion
disk. Therefore, it seems unlikely that the disk has been completely reformed,
although it may be in the process of reforming.

It is also possible that the He II emission is not associated with a reforming
accretion disk, but instead may originate in a separate structure of the
ejecta. \citet{2012arXiv1211.3112N}, suggest evidence of a late ejection of
material around 100 days after the outburst, which may explain the different
geometry of the He II lines or even the inner component of the [O III] lines.

In any case, the ejecta were not expanding with spherical symmetry even many
months after the X-ray grating observations. Assuming that they were partially
optically thick to soft X-rays while occupying only portion of the
line-of-sight to the WD seems a reasonable conclusion.

\section{DISCUSSION}
\label{conc}

Below we discuss five intriguing results of our analysis. 

\begin{enumerate}
\item X-ray emission lines originate in dense clumps (n$_{\rm e}\sim10^{10}$
  cm$^{-3}$) of cooling ejected plasma in CIE. The ``soft'' X-rays emission
  line spectrum of T Pyx bears similarities with CNe V1494 Aql
  \citep{2009AJ....137.4627R} and V382 Vel \citep{2005MNRAS.364.1015N} which
  showed strong emission lines over weak continuum at nearly the same time
  after the initial outburst as our T Pyx observations (268 days, V382 Vel;
  304 days, V1494 Aql). Both were modeled as a plasma in CIE with enhanced N
  abundances and had an absence of Fe line emission. While these three novae
  all share unevolved companions, their outbursts were quite different outside
  of the X-ray regime. V382 Vel and V1494 Aql are very fast novae with decay
  times (time to decline 3 mag from peak) of $t_3=10$ days and $t_3=16$ days
  \citep{2000A&A...355L...9K}, respectively. T Pyx, for reference, has a decay
  times of $t_3=62$ days \citep{2010ApJS..187..275S}.

  We would like to also suggest a possible interpretation, that the spectra of
  these three novae above may be due to internal shocks from the interaction
  of different velocity shells within the ejecta (see
  \citealt{2001ApJ...551.1024M}).

\item Many of the X-ray emission lines show blue-shifted centroids that are
  largest at higher ionization potentials. Comparing the X-ray lines to the
  {\sl SALT} nebular [O III] emission, which also trace ejected material from
  602 days after outburst, it is clear that both an approaching and receding
  outflow exists that is centered near zero velocity (see Figure
  \ref{fig:saltv}). At the time of our X-ray observations we note that the
  FWHM of the emission lines is much larger than their velocity offsets which
  we interpret as a partial absorption in the receding flow of the hardest
  X-rays lines observed.

\item The continuum flux in the {\sl Chandra} (day 210) X-ray spectra was best
  fit with a WD atmospheric model at a temperature of $\sim$420,000
  K. Assuming this was the temperature at the peak of the SSS-phase,
  theoretical models of nova outbursts for this atmospheric temperature
  predict a bolometric luminosity at, or above, $3\times 10^{37}$ ergs
  s$^{-1}$ \citep{2012BaltA..21...76S}. This is of course a lower limit given
  the fit temperature is likely less than at the outburst peak and indeed the
  nova may have turned off by the time of our observation. From our fit we
  derive $L_{Bol}$ to be a factor of 10 less than predicted. Furthermore, even
  at the peak of X-ray emission (a $\sim$6 times higher count rate, see Figure
  \ref{fig:slc}) the luminosity for the WD is still less than predicted.

  The low luminosity observed, along with evidence of primarily observing the
  face of the ejected material, may point to an obscuration of the WD flux.  A
  column density of N$_{\rm H}\sim4\times10^{21}$ cm$^{-2}$, or $\sim3$ times
  our best-fit value, would be required to match the expected flux at this
  temperature. This column corresponds to a ejected mass of a few times
  $10^{-6} M_\odot$ assuming a filling factor of 3\% and an outflow velocity
  of 4000 km s$^{-1}$ \citep{2011A&A...533L...8S}. This seems a reasonable
  value but of course requires ejected mass to be along the line-of-sight.

  It is also very puzzling that in the 25 days between X-ray grating
  observations, we observe no significant drop in the WD temperature but a
  factor of $>$2 decrease in continuum flux (see Figure \ref{fig:mspec}). All
  acceptable model fits show that N(H) does not seem to have increased
  drastically, which is what we expect at this stage. The rapid overall
  decrease in count rate (see Figure \ref{fig:slc}, Swift) is typical of the
  cooling curve, after nuclear burning has ceased, yet the WD does not appear
  to have cooled.
  
  We see two possibilities to interpret this phenomenon: First, that the
  atmospheric model we have used (\citealt{2010ApJ...717..363R}; hydrostatic
  atmosphere at a constant radius) may not be adequate to describe the X-ray
  spectra at this stage of the outburst. It is possible that the WD radius at
  the SSS peak was bloated due to a wind \citep[see][]{2012ApJ...756...43V},
  so that the initial decay in the X-ray light curve is due to the wind
  ceasing and the WD radius finally shrinking to typical WD dimensions. In
  other words, if Rauch's models are correct, we expect a cooling at constant
  radius, whereas if van Rossum's ``wind models'' should be adopted, we expect
  first, that the radius shrinks followed by a decrease in
  temperature. However, since van Rossum's models, so far, are only available
  with solar abundances (not typical of realistic novae) at this stage we do
  not repeat the fit to the Chandra spectrum with these models.

  Alternatively, \citet{2012arXiv1211.3112N} suggest a new episode of large,
  asymmetric mass ejection around day 100. Such material could decrease the
  flux while the WD temperature remained constant. We consider this
  possibility less likely at this late post-outburst stage, but it cannot be
  ruled out to explain what we observe in Figure \ref{fig:mspec} and the
  results of the atmospheric fit.

\item If the peak temperature did not vary significantly in the initial
  decline from the SSS phase, the short SSS turn-off time is unusual for an
  effective temperature T$_{\rm eff}$ of the order of 400,000 K
  \citep{2012BaltA..21...76S} (see examples of other novae in
  \citet{2012BASI...40..333O}, and references therein).  This may indicate
  that more unburned H-rich envelope is leftover here than in other novae.  If
  a large unburned envelope is present, this nova does not need to accrete a
  large quantity of material to trigger a new outburst. It may quickly reach
  the pressure for TNR within a few years, a cycle which will continue, as
  long as the TNR events are all somewhat ineffective in ejecting the accreted
  envelope. Such a temporary ``stunted'' cycle of RN eruptions every
  $\simeq$20 years may last until the WD manages to eject most of the of
  unprocessed accreted material.

  \citet{2010ApJ...708..381S} has suggested that an outburst prior to 1890 may
  have been the first after many centuries without eruption which, unlike the
  more recent events, was similar to a CN outburst. At that point T Pyx would
  initiate a phase of unsustainable irradiation induced mass transfer that,
  decaying over time, causes the present RN status to be transient in nature
  heading back toward an extended dormant state. If mass transfer was driven
  by an irradiated secondary, a high column density of the order of 10$^{22}$
  cm$^{-2}$ would have been necessary in the month before the 2011 outburst to
  shield the hot source. The low pre-outburst soft-X-ray flux, the short lived
  2011 SSS phase, and the low temperature observed in T Pyx indicate that
  there is not a prolonged irradiation induced mass transfer after each RN
  outburst.

  We would like to suggest a modification of the \citet{2010ApJ...708..381S}
  scenario, namely that the frequency of the outbursts and the short,
  under-luminous SSS duration indicate a nova that does not effectively expel
  its envelope in a single outburst. If T Pyx is an intermediate polar (IP),
  magnetic fields may be the cause of ineffective outbursts. Some analytical
  calculations were done by \citet{1988ApJ...330..264L} who, in the case of
  the stronger field of polars, find interference with convection, causing
  weaker outbursts than in a case with weak magnetic fields. On the other
  hand, a magnetic field may accelerate envelope expulsion once the TNR flash
  has occurred as in the magnetic rotator case analyzed by
  \citet{1992A&A...257..548O}, and thus ending a weak outburst quickly. For
  the time being this scenario is highly speculative, but we suggest that some
  unusual physical parameters are causing a long series of recurrent eruptions
  that do not eject the whole accreted envelope.

\item The orbital variability of the soft X-ray flux is not easy to interpret,
  but it is not unprecedented in CNe. It was observed in HV Cet
  \citep{2012A&A...545A.116B} and especially in V5116 Sgr
  \citep{2008ApJ...675L..93S} which for months showed a periodic decaying
  flare that \citet{2012BASI...40..333O} suggest may be due to resumed
  magnetic accretion that prevents the WD atmosphere from being thermally
  homogeneous. T Pyx may be a similar system in which renewed accretion
  streams to the polar caps prevent thermal homogeneity of the
  atmosphere. This conclusion is supported by the enhanced continuum flux of
  the elevated phase.

  If T Pyx has a magnetic WD, the concurrent ephemeris with the optical light
  curve could be interpreted as an IP. In this case, the measured period would
  not be orbital in nature but instead the WD rotation period. This would
  require the WD spin to be 1.8 hr making T Pyx the slowest rotating IP known
  \citep[beyond EX Hya at $\sim$1.2 hr,][]{1999MNRAS.310..203K}. For a
  ``disk-like'' truncated accretion disk the ratio of spin to orbital period
  would have to be less than a tenth, but there are possibilities of
  ``stream-like'' and ``ring-like'' accretion structures on the equatorial
  plane, where the true orbital period would not have to be much longer,
  possibly only by a factor of 40\% \citep{2008ApJ...672..524N}. The spectrum
  of the secondary should also otherwise be detected in quiescence, although
  in the case of CP Pup, \citet{2013arXiv1303.2777M} hypothesize that an
  ``accretion curtain'' may mask the secondary's spectrum and go undetected. A
  relatively strong magnetic field of an IP may also explain why this nova is
  different from most others.

  The above hypothesis is, however, at odds with the long-term increasing
  period observed by \citet{1998PASP..110..380P}, which would require the WD
  to be spinning down if it were an IP (opposite the expectation of an
  accreting WD). An alternative scenario could involve obscuring material
  blown off the secondary during the outburst or modulation of an accretion
  disk structure coming into the line-of-sight. This would imply the period we
  observe is indeed the orbital period. A partial obscuration scenario may
  however be at odds with the detection of the X-ray period in quiescence
  \citep{2012arXiv1208.6120B} and would require the inclination to be larger
  than claimed in the current literature. Both explanations remain highly
  speculative but seem reasonable interpretations of the data.
\end{enumerate}

\section{CONCLUSIONS}
\label{conclusion}

In this paper, we have analyzed spectral and temporal components of X-ray
grating and optical spectra during the decline of the 2011 outburst of RN T
Pyx. Below we summarize the main findings of this work.

\begin{enumerate}
\item The X-ray grating spectra of T Pyx during the decline of the super-soft
  phase show a significant contribution from emission lines to the X-ray
  flux. These lines appear to be emitted from high-density clumps
  (n$_e$$\sim$10$^{10}$ cm$^{-3}$) of ejected material in CIE. We interpret
  the blueshifted centroids of the lines as evidence of primarily seeing the
  face of the ejected material.

\item The X-ray continuum flux is modulated with commonly accepted orbital
  period.

\item We have not measured the peak T$_{\rm eff}$ of the WD and the nova may
  have in fact turned off before our observations. However, the fact that the
  temperature did not significantly vary between the {\sl Chandra} and the
  {\sl XMM-Newton} observations, while the overall continuum flux decreased,
  suggests that the initial decay occurred at an almost constant
  temperature. Either because, mass loss from a wind mechanism did not
  completely cease until weeks after the peak X-ray SSS luminosity, causing
  the initial decline to occur with a shrinking radius at near constant
  temperature. Or, because the a new ejection of mass increased the amount of
  absorbing material along the line-of-sight (this seems less likely given the
  late post-outburst stage).

\item The luminosity we derive is more than an order of magnitude lower than
  expected and even at the peak of the SSS flux, is at minimum a factor of
  $\sim$3 to small. The low luminosity along with mainly blue shifted X-ray
  signatures from the ejecta suggest the WD is partially obscured by the
  ejecta.

\item {\sl SALT} spectra display nearly symmetric bipolar ejecta almost 2 yr
  after the outburst revealed by double peaked emission lines of the [O III]
  doublet. He II 4686 \AA \ and H$\beta$ are also present but with different
  velocity structures either originating in a reforming accretion disk or a
  separate small episode of mass ejection.
\end{enumerate}
 
\acknowledgments The authors would like to thank Richard Townsend for his aid
in periodogram error analysis and Jay Gallagher for many useful
discussions. We would also like to acknowledge the use of public data from the
{\sl Swift} data archive and we acknowledge, with thanks, the variable star
observations from the AAVSO International Database contributed by observers
worldwide and used in this research. The Swift project at the University of
Leicester is supported by the UK Space Agency. This work has been supported by
a Chandra grant awarded by NASA through the Smithsonian Center for
Astrophysics.

\end{document}